\documentclass[twocolumn,english,aps,prb,floats]{revtex4-2}
\usepackage[T1]{fontenc}
\usepackage[latin9]{inputenc}
\usepackage{babel}
\usepackage{bm}
\usepackage{amsmath}
\usepackage{amssymb}
\usepackage{graphicx}
\usepackage{esint}
\usepackage[bookmarks=false,
 breaklinks=false,pdfborder={0 0 1},backref=false,colorlinks=false]
 {hyperref}

\makeatletter
\usepackage{babel}

\usepackage{bm}

\usepackage{siunitx}
\DeclareSIUnit\angstrom{\text{\AA}}
\usepackage{xcolor}

\@ifundefined{textcolor}{}{%
 \definecolor{BLACK}{gray}{0}
 \definecolor{WHITE}{gray}{1}
 \definecolor{RED}{rgb}{1,0,0}
 \definecolor{GREEN}{rgb}{0,1,0}
 \definecolor{BLUE}{rgb}{0,0,1}
 \definecolor{CYAN}{cmyk}{1,0,0,0}
 \definecolor{MAGENTA}{cmyk}{0,1,0,0}
 \definecolor{YELLOW}{cmyk}{0,0,1,0}
}

\usepackage{ifpdf}\usepackage{bm}

\makeatother

\begin{document}
\title{Oscillation modes of skyrmion strings in a ferromagnetic film}
\author{Eugene M. Chudnovsky and Dmitry A. Garanin}
\affiliation{Physics Department, Herbert H. Lehman College and Graduate School,
The City University of New York, 250 Bedford Park Boulevard West,
Bronx, New York 10468-1589, USA }
\date{\today}
\begin{abstract}
It is shown that a skyrmion string in a ferromagnetic film consisting
of $N$ atomic layers exhibits flexural oscillations at eigenfrequencies
$|\omega_{m}|=(4J'S/\hbar)\sin^{2}[\pi m/(2N)]$, where $J'$ is the
interlayer exchange coupling, $S$ is the length of the atomic spin,
and $m=1,...,N-1$. This result is confirmed numerically for individual
skyrmion strings in a discrete microscopic spin-lattice model of up
to ten atomic layers, as well as for skyrmion-string lattices at finite
temperature. 
\end{abstract}
\maketitle

\section{Introduction}

\label{Sec_Introduction}

Magnetic skyrmions have attracted considerable attention in recent
years due to their unique topology-driven physics and mathematics
\cite{Bogdanov-NatPhy2020}. Skyrmions have been observed in various
magnetic materials, where they can be stabilized by Dzyaloshinskii-Moriya
interaction (DMI) \cite{Bogdanov1989,Bogdanov94,Bogdanov-Nature2006,Heinze-Nature2011,Boulle-NatNano2016,Leonov-NJP2016},
frustrated exchange interactions \cite{Leonov-NatCom2015,Zhang-NatCom2017},
magnetic anisotropy \cite{IvanovPRB06,Lin-PRB2016}, disorder \cite{CG-NJP2018},
and geometrical confinement \cite{Moutafis-PRB2009}. They typically
emerge during changes in the magnetic field, when conventional laminar
magnetic domains transform into skyrmion lattices (SkL). Skyrmion-based
topologically protected information technology is being actively developed
\cite{Fert-NatMat2017,APL-2021}.

In recent years, excitation modes of individual skyrmions and SkL
have been studied \cite{Mochizuki-PRL2012,Onose-PRL2012,DA-RJ-EC-PRB2020,Aqeel-PRL2021,Satywali-NatCom2021,Lee-JPhys2022,Li-JPhys2023,DG-EC-PRB2025,EC-DG-PRB2026,DG-EC-MMM2026}.
Theoretical work in this field mostly treated skyrmions as 2D objects.
However, in a film of finite thickness, a magnetic skyrmion is actually
a stack of 2D skyrmions in each atomic layer, which should be treated
as a 3D skyrmion string. Current-induced dynamics of the SkL has been
studied experimentally by Yokouchi et al. \cite{Yokouchi-SviAdv2018},
who reported a frequency-dependent Hall effect, which they attributed
to the flexible nature of skyrmion strings. Seki et al. \cite{Seki-NatCom2020}
demonstrated the coherent propagation of spin excitations along skyrmion
strings, and provided direct visualization of skyrmion strings with
the help of magnetic X-ray tomography \cite{Seki-NatMat2022}. Kravchuk
studied the nonlinear dynamics of skyrmion strings \cite{Kravchuk-PRB2023},
by using micromagnetic simulations to compute helical waves along
the string. Bending of skyrmion strings under a thermal gradient has
been observed by Ran et al. \cite{Ran-NatCom2024}. Bending of skyrmion
strings under a thermal gradient has been observed by Ran et al. \cite{Ran-NatCom2024}.

\begin{figure}[h]
\centering{}\includegraphics[width=8cm]{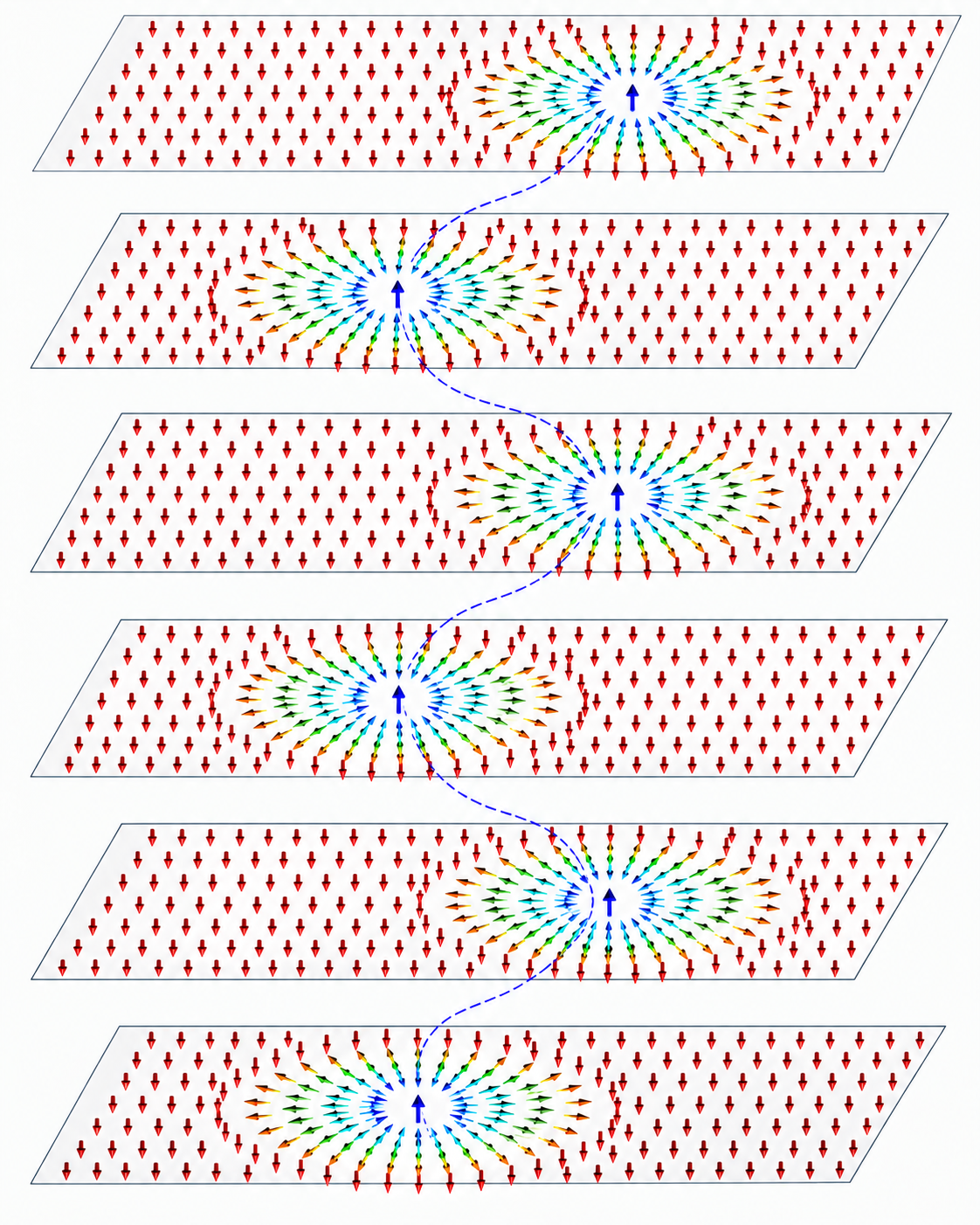} \caption{Conceptual illustration of a flexural eigenmode of a skyrmion string
extending through a ferromagnetic film. In practice, the amplitude
of lateral displacements of 2D skyrmions from the centerline of the
string would be small compared to the size of the skyrmion.}
\label{Fig_S-string} 
\end{figure}

Although string dynamics and propagating excitations have been studied
previously, to the best of our knowledge, an explicit analytical solution
for the flexural eigenmodes of a finite N-layer skyrmion string has
not been reported. In this paper we provide such a solution for the
flexural modes of the skyrmion string, see Fig.\ \ref{Fig_S-string}.
Our method is based upon the generalization of the method that was
applied to calculation a skyrmion mass in an antiferromagnet and a
two-component ferrimagnet \cite{Panigrahy2022,Nowak-PRB2023,Lau2025,EC-DG-PRB2026,DG-EC-MMM2026},
which considered displacements of the centers of skyrmions in the
two sublattices. We consider such displacements along the line of
the string in a ferromagnetic film consisting of $N$ atomic layers
and derive the frequencies of the flexural modes of the string, $|\omega_{m}|=(4J'S/\hbar)\sin^{2}[\pi m/(2N)]$,
where $J'$ is the interlayer exchange coupling, $S$ is the length
of the atomic spin, and $m=0,1,...,N-1$. We confirm this analytical
result numerically by computing the frequencies of the excitation
modes of the skyrmion string and skyrmion-string lattices in a discrete,
finite-temperature spin-lattice model considering up to ten atomic
layers.

The paper is organized as follows. Analytical theory is provided in
Section \ref{Sec_analytical}. Excitations induced by the displacement
of skyrmion centers in the adjacent layers of a two-layer ferromagnet
are studied in Subsection \ref{Sec_two-layers}. Subsection \ref{Sec_string}
generalizes the two-layer treatment for a film consisting of $N$
layers. It computes eigenmodes of flexural oscillations in a continuous
limit of $N\rightarrow\infty$, as well as in a discrete limit of
finite $N$. Oscillating dynamics generated by an oscillating force
is studied in Subsection \ref{Sec_dynamics}. The force produced by
a spin-orbit torque from a heavy metal substrate is calculated in
Subsection \ref{Sec_SO torque}. Numerical work is reported in Section
\ref{Sec_numerical_results}. The discrete spin-lattice model and
numerical method are formulated in Section \ref{Sec_Microscopic-model}.
Numerical results on the excitation modes of individual skyrmion strings
and lattices of skyrmion strings are reported in Subsection \ref{Sec_numerical_results}.
The results and their implications for experiments on skyrmion lattices
and microwave absorption by films of finite thickness that contain
skyrmions are discussed in Section \ref{Sec_discussion}.

\section{Analytical theory}

\label{Sec_analytical}

\subsection{Skyrmion in a two-layer ferromagnet}

\label{Sec_two-layers}

Consider first two ferromagnetic XY layers of spin densities ${\bm{\sigma}}_{1}({\bf r})$
and ${\bm{\sigma}}_{2}({\bf r}')$, interacting ferromagnetically,
with the exchange energy given by 
\begin{equation}
U_{12}=-\int d^{2}rd^{2}r'J({\bf r}-{\bf r}'){\bm{\sigma}}_{1}({\bf r})\cdot{\bm{\sigma}}_{2}({\bf r}').
\end{equation}
Assuming that the exchange interaction falls off exponentially at
$|{\bf r}-{\bf r}'|>a$, it makes sense to replace $J({\bf r}-{\bf r}')$
with $J'a^{2}\delta({\bf r}-{\bf r}')$, where $J'$ represents the
interlayer exchange constant. This gives $U_{12}=-J'a^{2}\int d^{2}r{\bm{\sigma}}_{1}({\bf r})\cdot{\bm{\sigma}}_{2}({\bf r})$.

For two 2D skyrmions of the same shape ${\bf f}({\bf r})$ (with ${\bf f}^{2}=1$)
centered at ${\bf r}={\bf R}_{1}$ in the first layer and at ${\bf r}={\bf R}_{2}$
in the second layer we can write ${\bm{\sigma}}_{1}=\sigma_{1}{\bf f}({\bf r}-{\bf R}_{1})$
and ${\bm{\sigma}}_{1}=\sigma_{2}{\bf f}({\bf r}-{\bf R}_{2})$, where
${\sigma}_{1,2}$ are constant spin densities of the uniformly magnetized
layers. This gives 
\begin{equation}
U_{12}=-J'a^{2}\sigma_{1}\sigma_{2}\int d^{2}r{\bf f}({\bf r})\cdot{\bf f}({\bf r}+{\bf d}),
\end{equation}
where ${\bf d}={\bf R}_{1}-{\bf R}_{2}$. When the displacement $d$
of skyrmions in the two layers with respect to each other is small
compared to their size, we can write 
\begin{equation}
{\bf f}({\bf r}+{\bf d})={\bf f}({\bf r})+d_{\alpha}\frac{\partial f}{\partial r_{\alpha}}+\frac{1}{2}d_{\alpha}d_{\beta}\frac{\partial^{2}f}{\partial r_{\alpha}\partial r_{\beta}}+...,
\end{equation}
where summation over $\alpha,\beta=x,y$ is implied. Substituting
this into the expression for $U_{12}$, integrating by parts, and
using the symmetry with respect to $x$ and $y$, we obtain 
\begin{equation}
U_{12}=\frac{1}{2}J'a^{2}\mathcal{D}\sigma_{1}\sigma_{2}{\bf d}^{2},\label{U12-d}
\end{equation}
where 
\begin{equation}
\mathcal{D}\equiv\frac{1}{2}\int dxdy\left[\left(\frac{\partial\mathbf{f}}{\partial x}\right)^{2}+\left(\frac{\partial\mathbf{f}}{\partial y}\right)^{2}\right].
\end{equation}
For a pure-exchange Belavin-Polyakov (BP) skyrmion \cite{BP} $\mathcal{D}=4\pi|Q|$,
with $Q$ being the topological charge.

The Thiele equations \cite{Thiele} describing skyrmion dynamics in
the two layers are 
\begin{equation}
-{\bf G}_{1}\times{\bf V}_{1}={\bf F}_{1}+{\bf F}_{12},\quad-{\bf G}_{2}\times{\bf V}_{2}={\bf F}_{2}+{\bf F}_{21},
\end{equation}
where ${\bf V}_{1,2}=\dot{{\bf R}}_{1,2}$ are velocities of the skyrmion
centers, ${\bf G}_{1,2}=4\pi\hbar Q{\sigma}_{1,2}\mathbf{e}_{z}$
are the gyrovectors, ${\bf F}_{1,2}$ are external forces acting on
the skyrmions in their respective layers, and ${\bf F}_{12}=-{\bf F}_{21}=-{\partial U_{12}}/{\partial{\bf R}_{1}}={\partial U_{12}}/{\partial{\bf R}_{2}}$.
Differentiating Eq.\ (\ref{U12-d}) with ${\bf d}={\bf R}_{1}-{\bf R}_{2}$,
we obtain ${\bf F}_{12}=-{\bf F}_{21}=-J'a^{2}\mathcal{D}\sigma_{1}\sigma_{2}{\bf d}$.
The corresponding Thiele equations can be written as 
\begin{eqnarray}
{\bf V}_{1} & = & \frac{1}{G_{1}}\mathbf{e}_{z}\times({\bf F}_{1}-J'a^{2}\mathcal{D}\sigma_{1}\sigma_{2}{\bf d})\\
{\bf V}_{2} & = & \frac{1}{G_{2}}\mathbf{e}_{z}\times({\bf F}_{2}+J'a^{2}\mathcal{D}\sigma_{1}\sigma_{2}{\bf d})
\end{eqnarray}
For the time derivative of the displacement, $\dot{{\bf d}}={\bf V}_{1}-{\bf V}_{2}$,
we obtain 
\begin{equation}
\dot{{\bf d}}=\frac{1}{G_{1}}\mathbf{e}_{z}\times{\bf F}_{1}-\frac{1}{G_{2}}\mathbf{e}_{z}\times{\bf F}_{2}-J'a^{2}\mathcal{D}\sigma_{1}\sigma_{2}\left(\frac{1}{G_{1}}+\frac{1}{G_{2}}\right)\mathbf{e}_{z}\times{\bf d}
\end{equation}
When the spin densities of the two layers are equal, it gives 
\begin{equation}
\dot{{\bf d}}=\frac{1}{G}\mathbf{e}_{z}\times({\bf F}_{1}-{\bf F}_{2})+\bm{\Omega}\times{\bf d},\label{d-dot}
\end{equation}
\begin{equation}
\bm{\Omega}=-\frac{2J'a^{2}\mathcal{D}\sigma^{2}}{G_{0}}\mathbf{e}_{z}=-\frac{2J'S\mathcal{D}}{4\pi\hbar Q}\mathbf{e}_{z},\label{Omega}
\end{equation}
where we used $\sigma={S}/{a^{2}},G_{0}=4\pi\hbar Q\sigma$. For a
BP skyrmion with $Q>0$, 
\begin{equation}
{\bm{\Omega}}=-\frac{2J'S}{\hbar}\mathbf{e}_{z}.\label{Omega-BP}
\end{equation}
When the forces ${\bf F}_{1,2}$ are zero or equal, Eq.\ (\ref{d-dot})
describes the skyrmions in the two layers circling each other around
a stationary center at the angular velocity ${\bm{\Omega}}$. The
sign in Eq.\ (\ref{Omega}) shows the direction of rotation.

\subsection{Skyrmion string in a ferromagnetic film}

\label{Sec_string}

Consider now flexural oscillations of a skyrmion string in the z-direction
passing through $N$ layers, see Fig.\ \ref{Fig_S-string}. The interlayer
energy is 
\begin{equation}
U=\frac{\mathcal{T}}{2}\sum_{n}^{N-1}|{\bf R}_{n}-{\bf R}_{n+1}|^{2}=\frac{\mathcal{T}}{2}\sum_{n}^{N-1}|{\bf d}_{n}|^{2},
\end{equation}
with 
\begin{equation}
\mathcal{T}\equiv J'a^{2}\mathcal{D}\sigma^{2}
\end{equation}
playing the role of tension. For interior layers, $2\leq n\leq N-1$,
the interaction forces are given by
\begin{equation}
-\frac{\partial U}{\partial{\bf R}_{n}}=\mathcal{T}({\bf R}_{n-1}-2{\bf R}_{n}+{\bf R}_{n+1}).
\end{equation}
At the free surfaces, 
\begin{equation}
-\frac{\partial U}{\partial{\bf R}_{1}}=\mathcal{T}({\bf R}_{2}-{\bf R}_{1}),\quad-\frac{\partial U}{\partial{\bf R}_{N}}=\mathcal{T}({\bf R}_{N-1}-{\bf R}_{N}).
\end{equation}
Thus, the Thiele equations for the individual skyrmion centers for
$1\leq n\leq N$ are 
\begin{equation}
\dot{{\bf R}}_{n}=\frac{1}{G_{0}}{\bf e}_{z}\times\left({\bf F}_{n}-\frac{\partial U}{\partial{\bf R}_{n}}\right),
\end{equation}
where ${\bf F}_{n}$ are external forces. Subtracting the equation
for ${\bf R}_{n}$ and ${\bf R}_{n+1}$ gives the string generalization
of Eq.\ (\ref{d-dot}) for $d_{n}(t)={\bf R}_{n}-{\bf R_{n+1}}$:
\begin{equation}
\dot{d}_{n}=\frac{1}{G_{0}}\mathbf{e}_{z}\times[{\bf F}_{n}-{\bf F}_{n+1}+\mathcal{T}({\bf d}_{n-1}-2{\bf d}_{n}+{\bf d}_{n+1})].\label{F_n}
\end{equation}
The boundary convention ${\bf d}_{0}={\bf d}_{N}=0$ allows one equation
to be used for all $n=1,...,N$. For $N=2$, only ${\bf d}_{1}$ exists,
which recovers Eq.\ (\ref{d-dot}).

The eigenfunctions $\phi_{n}$ and eigenvalues $\lambda_{n}$ of the
discrete Laplacian \cite{Strang-1999} satisfy 
\begin{equation}
\phi_{n-1}^{(m)}-2\phi_{n}^{(m)}+\phi_{n+1}^{(m)}=-\lambda_{m}\phi_{n}^{(m)}
\end{equation}
with $\phi_{0}^{(m)}=\phi_{N}^{(m)}=0$, which gives 
\begin{equation}
\phi_{n}{(m)}=\sqrt{\frac{2}{N}}\sin\left(\frac{\pi mn}{N}\right),\quad\lambda_{m}=4\sin^{2}\left(\frac{\pi m}{2N}\right).
\end{equation}

Substituting ${\bf d}_{n}\propto e^{-i\omega t}\phi_{n}$ into Eq.\ (\ref{F_n},
we obtain for the frequencies of flexural oscillations 
\begin{equation}
\Omega_{m}=-\frac{\mathcal{T}}{G_{0}}\lambda_{m}=-\frac{4\mathcal{T}}{G_{0}}\sin^{2}\left(\frac{\pi m}{2N}\right).
\end{equation}
For a BP skyrmion string it gives 
\begin{equation}
\omega_{m}=-\frac{4J'S}{\hbar}\sin^{2}\left(\frac{\pi m}{2N}\right).\label{string-general}
\end{equation}

In a bilayer case of $N=2$, there are two modes: $m=0$ and $m=1$.
The $m=0$ mode corresponds to the rigid translation of the bilayer
skyrmion. The $m=m_{1}$ mode corresponds to the bilayer result, Eq.\ (\ref{Omega-BP}),
of twin skyrmions circling each other at the angular frequency 
\begin{equation}
\omega_{1}=-\frac{4J'S}{\hbar}\sin^{2}\left(\frac{\pi}{4}\right)=-\frac{2J'S}{\hbar}.
\end{equation}

The continuous limit of a film of thickness $b$ with $N\rightarrow\infty$,
one has $d_{n\pm1}-d_{n}\rightarrow\pm b\partial_{z}d+\frac{1}{2}b^{2}\partial_{z}^{2}d$,
leading to $d_{n-1}-2d_{n}+d_{n+1}\rightarrow b^{2}\partial_{z}^{2}d$.
In that limit we obtain 
\begin{equation}
\partial_{t}{\bf d}(z,t)=\frac{1}{G_{0}}\mathbf{e}_{z}\times[b\partial_{z}{\bf F}(z,t)+\mathcal{T}b^{2}\partial_{z}^{2}{\bf d}(z,t)]
\end{equation}
The equation for the mode ${\bf d}(z,t)={\bf d}_{k}(t)\sin(kz)$ is
\begin{equation}
\dot{{\bf d}}_{k}=-\frac{\mathcal{T}b^{2}k^{2}}{G_{0}}\mathbf{e}_{z}\times{\bf d}_{k}+\frac{1}{G_{0}}\mathbf{e}_{z}\times{\bf F}_{k},
\end{equation}
where ${\bf F}_{k}$ is the Fourier component of the force difference
in Eq.\ (\ref{F_n}). Equivalently, 
\begin{equation}
\dot{{\bf d}}_{k}=\Omega_{k}\mathbf{e}_{z}\times{\bf d}_{k}+\frac{1}{G_{0}}\mathbf{e}_{z}\times{\bf F}_{k},\label{k-mode}
\end{equation}
where 
\begin{equation}
\Omega_{k}=-\frac{J'a^{2}\mathcal{D}\sigma}{4\pi\hbar Q}(bk)^{2}
\end{equation}
is the angular frequency of the flexural mode of the skyrmion string
with a wave vector $k$. For a BP skyrmion with $\mathcal{D}=4\pi|Q|$,
substituting here $\sigma=S/a^{2}$, we obtain 
\begin{equation}
\omega_{k}=-\frac{J'S}{\hbar}(bk)^{2},
\end{equation}
which coincides with Eq. (\ref{string-general}) in the limit of small
$m$ and $k_{m}=m\pi/\hbar$.

\subsection{Driven oscillations of the skyrmion string}

\label{Sec_dynamics}

Introducing ${\bf f}_{k}\equiv{\bf F}_{k}/G_{0}$ into the full equation
(\ref{k-mode}) for the $k$-mode, and choosing ${\bf f}_{k}=f_{k}\mathbf{e}_{x}$,
we obtain 
\begin{equation}
\dot{d}_{kx}=-\Omega_{k}d_{ky},\qquad\dot{d}_{ky}=\Omega_{k}d_{kx}+f_{k}
\end{equation}
or 
\begin{equation}
\ddot{d}_{kx}+\Omega_{k}^{2}d_{kx}=-\Omega_{k}f_{k},\qquad d_{ky}=-\frac{1}{\Omega_{k}}\dot{d}_{kx}.
\end{equation}
The most interesting case is the one of the external force oscillating
in resonance with $k$-mode, $f_{k}=f\cos(\Omega_{k}t)$. The general
solution with two arbitrary constants $A$ and $\alpha$ is 
\begin{eqnarray}
d_{kx} & = & A\cos(\Omega_{k}t+\alpha)-\frac{1}{2}ft\sin(\Omega_{k}t)\\
d_{ky} & = & A\sin(\Omega_{k}t+\alpha)+\frac{1}{2\Omega_{k}}f\sin(\Omega_{k}t)+\frac{1}{2}ft\cos(\Omega_{k}t)\nonumber \\
\end{eqnarray}
At large times it is dominated by terms linear in time, making the
amplitude $d_{k}$ of the skyrmion string oscillations proportional
to $t$: $d_{k}=\sqrt{d_{kx}^{2}+d_{ky}^{2}}=\frac{1}{2}|f|t$. The
energy pumped into the flexural mode by a resonant spin current or
microwaves will grow as $U_{12}\propto{\bf d}_{k}^{2}\propto t^{2}$.
This would eventually drive the oscillation amplitude beyond the regime
of validity of the linear approximation and potentially destabilize
the skyrmion string. In practice, however, the linear growth of the
amplitude of flexural oscillations will be limited by dissipation.
To account for dissipation, we write Eq.\ (\ref{k-mode}) in the
form 
\begin{equation}
\dot{{\bf d}}_{k}=\Omega_{k}\mathbf{e}_{z}\times{\bf d}_{k}-\Gamma_{k}{\bf d}_{k}+\frac{1}{G_{0}}\mathbf{e}_{z}\times{\bf F}_{k},\label{d-dot-dis}
\end{equation}
where $\Gamma_{k}$ is the damping rate of the k-th flexural mode.
At $f_{k}=f\cos(\Omega_{k}t)$, the asymptotic solution for weak damping,
$\Gamma_{k}\ll\Omega_{k}$, is $d_{kx}=-[{f}/({2\Gamma_{k}})]\sin(\Omega_{k}t),d_{ky}=[{f}/({2\Gamma_{k}})]\cos(\Omega_{k}t)$,
so that $d_{k}=\sqrt{d_{kx}^{2}+d_{ky}^{2}}={|f|}/({2\Gamma_{k}})$.

From a practical perspective, it may be easier to excite flexural
oscillations of the skyrmion string by the time-dependent force applied
to the bottom layer only, e.g., induced by the spin current in the
substrate on which the film is deposited. This would correspond to
${\bf F}_{n}(t)={\bf F}(t)\delta_{n1}$. Replacing $k$ with $n$
in Eq.\ (\ref{k-mode}), expanding ${\bf d}_{n}$ and ${\bf F}_{n}-{\bf F}_{n+1}$
in terms of the normal-mode functions $\sin(\pi mn/N)$ as 
\begin{eqnarray}
 &  & {\bf d}_{n}=\sum_{m=1}^{N-1}{\bf d}_{m}(t)\sin\left(\frac{\pi mn}{N}\right)\label{dn-dm}\\
 &  & {\bf F}_{n}-{\bf F}_{n+1}=\sum_{m=1}^{N-1}{\bf F}_{m}(t)\sin\left(\frac{\pi mn}{N}\right),
\end{eqnarray}
and using the orthogonality condition 
\begin{equation}
\sum_{n=1}^{N-1}\sin\left(\frac{\pi mn}{N}\right)\sin\left(\frac{\pi m'n}{N}\right)=\frac{N}{2}\delta_{mm'},
\end{equation}
we obtain 
\begin{eqnarray}
{\bf F}_{m}(t) & = & \frac{2}{N}\sum_{n=1}^{N-1}({\bf F}_{n}-{\bf F}_{n+1})\sin\left(\frac{\pi mn}{N}\right)\nonumber \\
 & = & \frac{2}{N}{\bf F}(t)\sin\left(\frac{\pi m}{N}\right).
\end{eqnarray}
\begin{equation}
\dot{{\bf d}}_{m}=\Omega_{m}\mathbf{e}_{z}\times{\bf d}_{m}+\frac{2}{NG_{0}}\sin\left(\frac{\pi m}{N}\right)\mathbf{e}_{z}\times{\bf F}(t),\label{d_m-F}
\end{equation}
where 
\begin{equation}
\Omega_{m}=-\frac{4\mathcal{T}}{G_{0}}\sin^{2}\left(\frac{\pi m}{2N}\right).
\end{equation}

For a harmonic force ${\bf F}(t)=F\cos(\omega t)\mathbf{e}_{z}$,
the effective driving force in Eq.\ (\ref{d_m-F}) is 
\begin{equation}
f_{m}(t)=\frac{2F}{NG_{0}}\sin\left(\frac{\pi m}{N}\right)\cos(\omega t).
\end{equation}
The physics is the same as driving oscillations of a string by shaking
one of its ends. Since $\sin(\pi m/N)\neq0$ for all $m=1,2,...,N-1$,
the force applied to the bottom layer excites every flexural mode.
However, the resonant excitation only occurs when its angular frequency
$\omega$ satisfies $\omega\approx|\Omega_{m}|$. Without damping,
the amplitude of the flexural mode at $\omega=\Omega_{m}$ grows with
time as 
\begin{equation}
d_{m}(t)=\frac{|F|}{NG_{0}}\sin\left(\frac{\pi m}{N}\right)t.
\end{equation}
With damping $\Gamma_{m}$, the amplitude saturates at 
\begin{equation}
d_{m}^{res}=\frac{|F|}{NG_{0}\Gamma_{m}}\sin\left(\frac{\pi m}{N}\right).
\end{equation}
\\

\subsection{Exciting flexural oscillations of the skyrmion string by a spin-orbit
torque from a heavy metal substrate}

\label{Sec_SO torque}

For a spin-orbit torque generated by a charge-current density ${\bf j}(t)$
in the substrate, the force acting on the skyrmion in the bottom layer
can be written as \cite{Gobel-2021,Yang-2023} 
\begin{equation}
F_{i}(t)=-\frac{\hbar\theta_{SH}\xi}{2e}I_{ij}[\mathbf{e}_{n}\times{\bf j}(t)]_{j}.
\end{equation}
where $\theta_{SH}$ is the spin Hall angle of the spin-source layer
(usually a heavy metal), $0<\xi<1$ is the spin-transmission factor,
$e>0$ is the elementary charge, $\mathbf{e}_{n}$ is the unit vector
normal to the interface, 
\begin{equation}
I_{ij}=\int d^{2}r[\partial_{i}{\bf f}({\bf r})\times{\bf f}({\bf r})]_{j},
\end{equation}
and ${\bf f}({\bf r})$ is the unit-vector magnetization of the skyrmion.
If ${\bf n}=\mathbf{e}_{z}$ and ${\bf j}=j(t)\mathbf{e}_{x}$, this
gives 
\begin{eqnarray}
F_{x} & = & -\frac{\hbar\theta_{SH}\xi}{2e}I_{xy}j(t)\\
F_{y} & = & -\frac{\hbar\theta_{SH}\xi}{2e}I_{yy}j(t).
\end{eqnarray}

For an axisymmetric Néel skyrmion with ${\bf m}(r,\phi)=\sin\theta(r){\bf e}_{r}+\cos\theta(r)\mathbf{e}_{z}$,
the nonzero components of tensor $I_{ij}$ are $I_{xy}=-I_{yx}\equiv I_{N}$,
$I_{xx}=I_{yy}=0$, where 
\begin{equation}
I_{N}=\pi\int_{0}^{\infty}dr\left[r\frac{d\theta}{dr}+\sin\theta\cos\theta\right].
\end{equation}
The BP skyrmion has $\theta({\bf r})=2\tan^{-1}(\lambda/r)$, with
$\lambda$ characterizing the size of the skyrmion. This gives $I_{N}=-2\pi\lambda$,
and the only nonzero component of the force becomes 
\begin{equation}
F_{x}=\frac{\pi\hbar\theta_{SH}\xi\lambda}{e}j(t)\label{F-j}
\end{equation}

Flexural oscillations of the skyrmion string must be excited by the
current in the substrate $j(t)$ oscillating at a resonance frequency
as described in the previous subsection. The energy of the $m$-th
mode is $U_{m}=N\mathcal{T}|{\bf d}_{m}|^{2}/4$. From Eq.\ (\ref{d-dot-dis})
the damping contribution to the energy derivative is 
\begin{equation}
\left(\frac{dU_{m}}{dt}\right)_{diss}=-\frac{N\mathcal{T}}{2}\Gamma_{m}|{\bf d}_{m}|^{2}.
\end{equation}
Consequently, in the steady state, the cycle-averaged absorbed power
is 
\begin{equation}
\bar{P}_{m}=\frac{N\mathcal{T}}{2}\Gamma_{m}\langle|{\bf d}_{m}(t)|^{2}\rangle.
\end{equation}
For a linearly oscillating force, $F(t)=F_{0}\cos(\omega t)$, the
steady state solution of Eq.\ (\ref{d-dot-dis}) gives 
\begin{equation}
\langle|{\bf d}_{m}|^{2}\rangle=\frac{f_{m}^{2}}{4}\left[\frac{1}{(\omega-|\Omega_{m}|)^{2}+\Gamma_{m}^{2}}+\frac{1}{(\omega+|\Omega_{m}|)^{2}+\Gamma_{m}^{2}}\right],
\end{equation}
where $f_{m}=[2F/(NG_{0})]\sin(\pi m/N)$.

For $j(t)=j_{0}\cos(\omega t)$, expressing $F_{0}$ via $j_{0}$
according to Eq.\ (\ref{F-j}), we obtain for a positive-frequency
resonance 
\begin{equation}
\bar{P}_{m}(\omega)=\frac{\mathcal{T}\Gamma_{m}}{2NG_{0}^{2}}\left(\frac{\pi\hbar\theta_{SH}\xi\lambda}{e}\right)^{2}\frac{j_{0}^{2}\sin^{2}({\pi m}/{N})}{(\omega-|\Omega_{m}|)^{2}+\Gamma_{m}^{2}}.
\end{equation}
Notice that since the string tension $\mathcal{T}$ is proportional
to $\sigma^{2}$, and the gyro factor $G_{0}$ is proportional to
$\sigma$, the spin density $\sigma$ cancels out from the expression
for the power. At resonance, for a $Q=1$ BP skyrmion 
\begin{equation}
\bar{P}_{m}^{res}(\omega)=\frac{\pi J'a^{2}\theta_{SH}^{2}\xi^{2}\lambda^{2}}{8e^{2}N\Gamma_{m}}j_{0}^{2}\sin^{2}\left(\frac{\pi m}{N}\right).
\end{equation}
The total power absorption by one string is 
\begin{equation}
\bar{P}_{{\rm string}}(\omega)=\sum_{m=1}^{N-1}\bar{P}_{m}(\omega).
\end{equation}
For a skyrmion lattice, this expression must be multiplied by the
areal density of the strings, $n_{sk}\sim1/\lambda^{2}$, to obtain
the estimate of the absorbed power per unit area of the film, which
makes it weakly dependent on the skyrmion size $\lambda$. \\

\section{Microscopic model and numerical approach}

\subsection{Microscopic model and numerical methods}

\label{Sec_Microscopic-model}

To test our analytical theory, we studied numerically the flexural
excitations in a 3D ferromagnetic model on a simple cubic lattice
with the Dzyaloshinskii-Moriya interaction (DMI) that supports skyrmions
in $x,y$ layers. This model is described by the Hamiltonian 
\begin{eqnarray}
{\cal H} & = & -\frac{1}{2}\sum_{ij}J_{ij}{\bf s}_{i}\cdot{\bf s}_{j}-\frac{D}{2}\sum_{i}s_{i,z}^{2}-\mathrm{H}\cdot\sum_{i}\mathbf{s}_{i}\nonumber \\
 & + & A\sum_{i}\left[\left({\bf s}_{i}\times{\bf s}_{i+\delta_{x}}\right){}_{x}+\left({\bf s}_{i}\times{\bf s}_{i+\delta_{y}}\right){}_{y}\right].\label{Ham}
\end{eqnarray}
Here $J_{ij}$ is the ferromagnetic nearest-neighbor exchange between
the unit-length classical-vector spins ${\bf s}_{i}$ with the coupling
constants $J>0$ within the $x,y$ layers and $J'>0$ between the
layers. Other terms are the easy-axis anisotropy and the Zeeman term,
as well as the DMI. In the DMI term, the subscript $i+\delta_{x}$stands
for the nearest neighbor of site $i$ in the positive $x$ direction.
We choose $J'/J=0.2$, $A/J=0.2,$and $H/J=-0.03$ as our main set
of parameters. To the Hamiltonian above, one can add the time-dependent
microwave (MW) field.

The dynamics of the lattice spins is described by the Landau-Lifshitz
(LL) equation, augmented by the spin-current term: 
\begin{equation}
\hbar\frac{\partial\mathbf{s}_{i}}{\partial t}+\hbar\left(\mathbf{v}_{S}\cdot\nabla\right)\mathbf{s}_{i}=\mathbf{s}_{i}\times\mathbf{H}_{\mathrm{eff},i},\label{LL}
\end{equation}
where $\mathbf{H}_{\mathrm{eff},i}=-\partial\mathcal{H}/\partial\mathbf{s}_{i}$
is the effective field in the energy units and $\mathbf{v}_{S}$ is
the spin current term. The action of this streaming term is translating
the whole set of $\mathbf{s}_{i}$ with the velocity $\mathbf{v}_{S}$.
To excite flexural modes, we use the spin current in the boundary
layer only. We do not include the phenomenological damping in the
LL equation, since the model shows a significant intrinsic damping
due to thermal effects.

In computations, we used, as usual, $J\Rightarrow1$, $S\Rightarrow1$,
and $\hbar\Rightarrow1$, as well as $a=b\Rightarrow1$. To find the
equilibrium states of the system with a single skyrmion (SS) or a
skyrmion lattice (SkL) at $T=0$, we performed the energy minimization
(see, e.g., Ref. \cite{gar25jpcm}) starting from the appropriate
initial conditions having the required value of the topological charge
$Q$. To find the flexural modes' frequencies, we excited the system
with the sinc spin current in the boundary (first) layer, $\mathbf{v}_{S}(t)=\mathbf{e}_{x}v_{S,0}\sin\left(\omega_{\max}t\right)/\left(\omega_{\max}t\right)$.
The Fourier spectrum of this function is a constant up to the cutoff
frequency $\omega_{\max}$ and zero above it, so that it excites all
modes in the interval $0<\omega<\omega_{\max}$. The LL equation was
solved with the 5th-order Butcher's Runge-Kutta method (RK5) that
makes 6 function evaluations per one integration step. We used moderate
integration times such as $t_{\max}\approx10000$ and computed the
fluctuation spectrum (FS) of the quantity $F$ as follows: 
\begin{equation}
\mathrm{FS}(\omega)=\frac{1}{t_{\max}}\left|\intop_{0}^{^{t_{\max}}}f(t)e^{i\omega t}dt\right|^{2}.\label{FS-2}
\end{equation}
As the quantity $f$, one can use the $X$ displacement of the skyrmion
(or the average displacement of skyrmions in the SkL) in the first
layer, $X$, from its equilibrium position. The modes' frequencies
were searched for as the resonances in $\mathrm{FS}(\omega)$. The
position of the skyrmion's center $\mathbf{R}$ within each layer
was defined with the help of the skyrmion-locator formula \cite{garchu24jmmm}
\begin{equation}
\mathbf{R}=\left.\sum_{s_{i,z}>0}\mathbf{r}_{i}s_{i,z}^{2}\right/\sum_{s_{i,z}>0}s_{i,z}^{2},\label{Skyrmion_locator}
\end{equation}
where the summation is performed within the layer and $i\in\mathrm{Skyrmion}$
are all lattice sites that belong to the skyrmion. Here the weight
factor $s_{z,i}^{2}$ favors the sites closer to the skyrmion's top.
The same formula was used for the SkL to describe the motion of all
skyrmions.

The analytical theory of Sec. \ref{Sec_string} is formulated in terms
of the relative displacements ${\bf d}_{n}={\bf R}_{n}-{\bf R}_{n-1}$,
whose normal modes are proportional to $\sin(\pi mn/N)$. For numerical
work, it is more convenient to use directly the skyrmion positions
${\bf R}_{n}$. The corresponding eigenvectors of the discrete Laplacian
with free boundaries are $\cos[(\pi m/N)(n-1/2)]$ (see Eq. (4.46)
of Ref. \cite{kacgar2001PA_surf}), providing 
\begin{equation}
{\bf R}_{n}(t)=\bar{{\bf R}}(t)+\sum_{m=1}^{N-1}{\bf R}_{m}(t)\cos\left[\frac{\pi m}{N}\left(n-\frac{1}{2}\right)\right].\label{R_n}
\end{equation}
Here $\bar{{\bf R}}=N^{-1}\sum_{n=1}^{N}{\bf R}_{n}(t)$ corresponds
to the $m=0$ cosine mode that describes the average position of the
skyrmion string. The relation to the sine eigenfunctions is through
\begin{equation}
{\bf d}_{n}^{(m)}=2\sin\left(\frac{\pi m}{2N}\right){\bf R}_{m}(t)\sin\left(\frac{\pi mn}{N}\right),
\end{equation}
where 
\begin{equation}
{\bf R}_{m}(t)=\frac{2}{N}\sum_{n=1}^{N}[{\bf R}_{n}(t)-\bar{{\bf R}}(t)]\cos\left[\frac{\pi m}{N}\left(n-\frac{1}{2}\right)\right],\label{Rm_def}
\end{equation}
Therefore, both representations describe the same flexural modes.
For numerical work, one can take the $X$ component of any ${\bf R}_{m}(t)$
and use it in Eq. (\ref{FS-2}) as $f(t)$.

The skyrmion size $\lambda_{\mathrm{eff}}$ was computed as \cite{caichugar2012prb}
\begin{equation}
\lambda_{\mathrm{eff}}^{2}=\frac{n-1}{2^{n}\pi N_{S}}\sum_{i}\left(1+s_{i,z}\right)^{n}\label{lam_eff_def}
\end{equation}
with summation performed within the layer and $n=4$. Here $N_{S}$
is the number of skyrmions in the system. This formula yields the
exact result for the skyrmion size $\lambda$ in the case of the pure-exchange
BP skyrmion.

Another type of numerical experiment is computing the FS in a thermal
state at a fixed temperature. Such thermal states are created by Metropolis
Monte Carlo (see, e.g., Ref. \cite{gar25jpcm}). We define the topological
charge for discrete spins on the lattice using the formula for the
body angle circumscribed by a triad of vectors \cite{eriksson90mm}.
The result has the form 
\begin{equation}
Q=\frac{1}{2\pi}\sum_{i,\epsilon=\pm1}\arctan\frac{\mathbf{s}_{i}\cdot\left(\mathbf{s}_{j}\times\mathbf{s}_{k}\right)}{1+\mathbf{s}_{i}\cdot\mathbf{s}_{j}+\mathbf{s}_{j}\cdot\mathbf{s}_{k}+\mathbf{s}_{k}\cdot\mathbf{s}_{i}},\label{Q_def_discrete}
\end{equation}
where $j\equiv i+\epsilon\delta_{x}$ and $k\equiv i+\epsilon\delta_{y}$.
This formula works well even at elevated temperatures where the deviation
of neighboring spins from each other is not small and the standard
formula for $Q$ using derivatives fails.

Running dynamical evolution for longer times leads to energy drift
and thus requires an energy correction that is performed in time intervals
as described in Refs. \cite{gar21pre,DG-EC-PRB2025}. To correct the
energy, we use the spin temperature $T_{s}$ as the control variable,
as done in Ref. \cite{DG-EC-PRB2025}. In the absence of single-site
interactions $T_{s}$ is given by 
\begin{equation}
T_{S}\equiv\frac{1}{2}\frac{\sum_{i}\left(\mathbf{s}_{i}\times\mathbf{H}_{\mathrm{eff},i}\right)^{2}}{\sum_{i}\mathbf{s}_{i}\cdot\mathbf{H}_{\mathrm{eff},i}},\label{TS}
\end{equation}
where $\mathbf{H}_{\mathrm{eff},i}$ is the effective field defined
in Eq. (\ref{LL}). Computation with the energy correction yields
a true stationary process in which spin energy and spin temperature
only weakly fluctuate around their thermodynamic values.

For the system with SkL, we used the system size $116\times132\times N$
in lattice units with periodic boundary conditions in the $x$ and
$y$ directions and free boundary conditions in the $z$ direction,
perpendicular to the film. For the film thickness (the number of layers),
we choose $N=10$. The layers of the size $116\times132$ neatly comprise
an SkL of 12 skyrmion strings directed along $z$. For the system
with a single skyrmion string, we used smaller system sizes, such
as $50\times50\times10$. The computations were performed with \textsc{Wolfram
Mathematica} with compilation, vectorization, and parallelization.

\subsection{Numerical results}

\label{Sec_numerical_results}

\begin{figure}[h]
\centering{}\includegraphics[width=8cm]{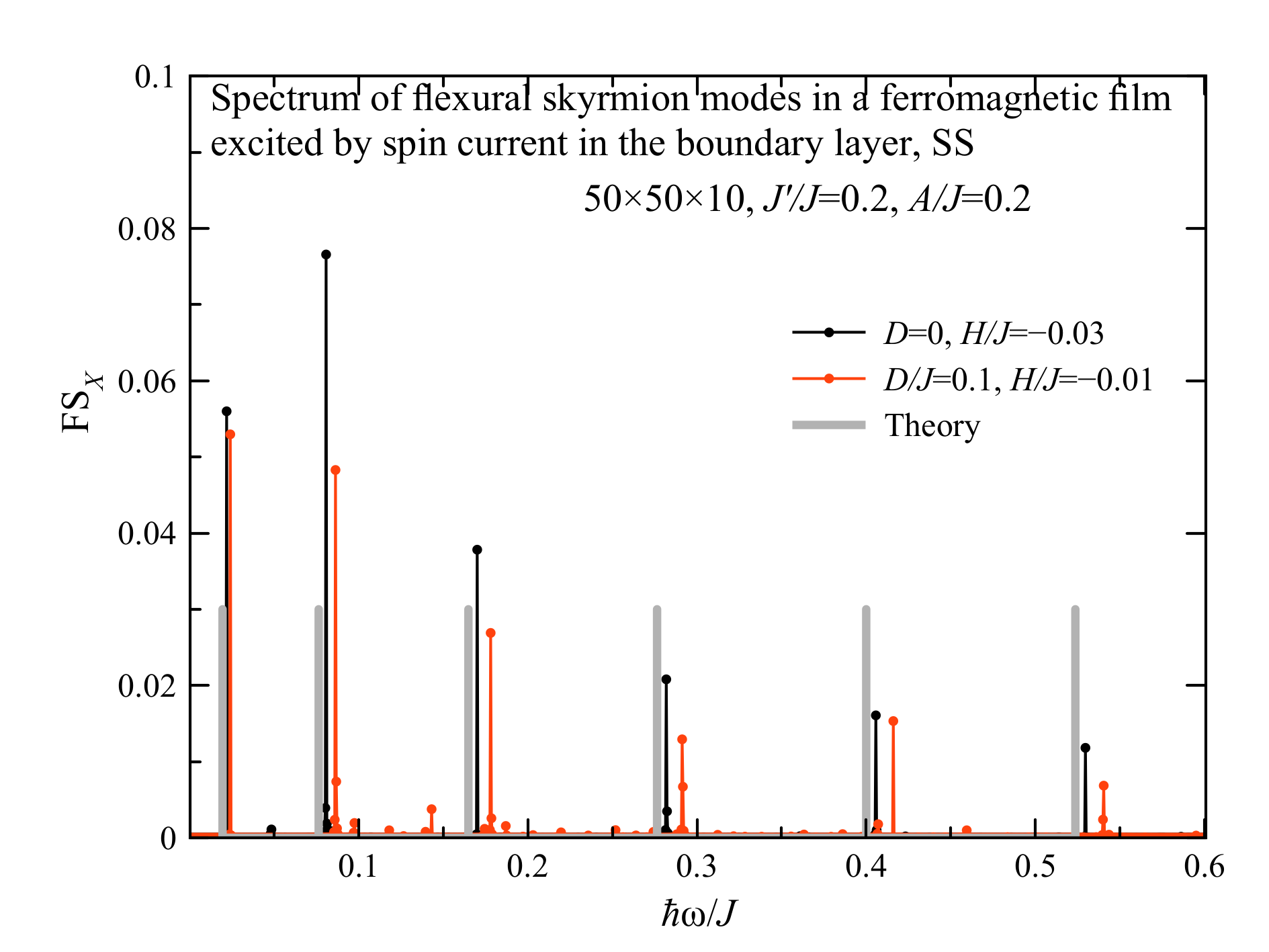}
\caption{Flexural resonances with $m=1,...,6$ of a single skyrmion string,
excited by the spin current at the boundary of a ferromagnetic film,
showing in the fluctuation spectrum of the $X$ displacement of the
skyrmion in the boundary layer. Gray vertical lines show the theoretical
result, Eq. (\ref{string-general}). }
\label{Fig_FSX_SS} 
\end{figure}

\begin{figure}[h]
\centering{}\includegraphics[width=8cm]{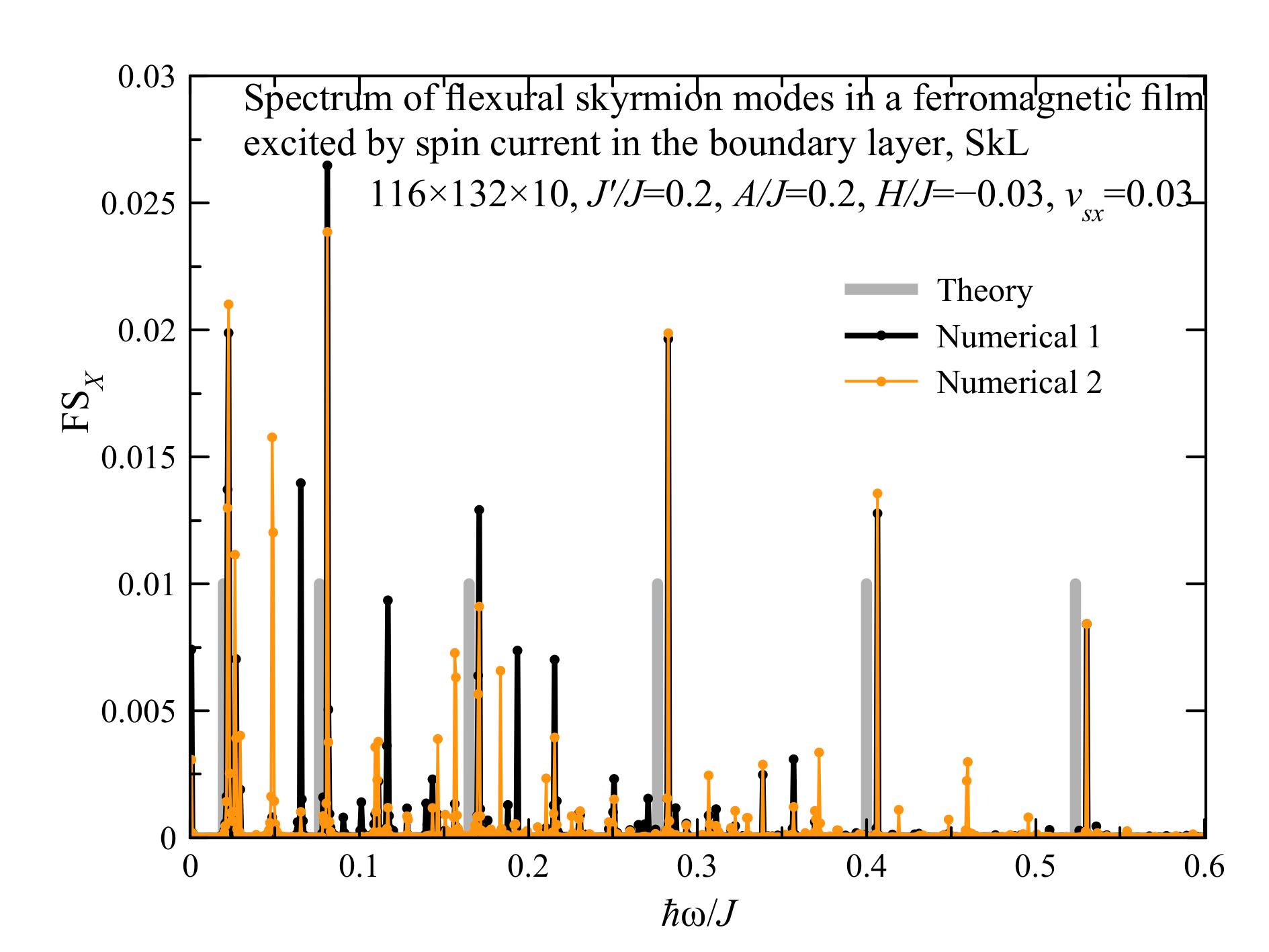}
\caption{The same for a larger system with an SkL containing 12 skyrmion strings.
Computations performed on two different computers show peaks in accordance
with the theory and with each other, as well as additional non-coinciding
peaks that can be interpreted as noise. }
\label{Fig_FSx-SkL} 
\end{figure}

\begin{figure}
\begin{centering}
\includegraphics[width=8cm]{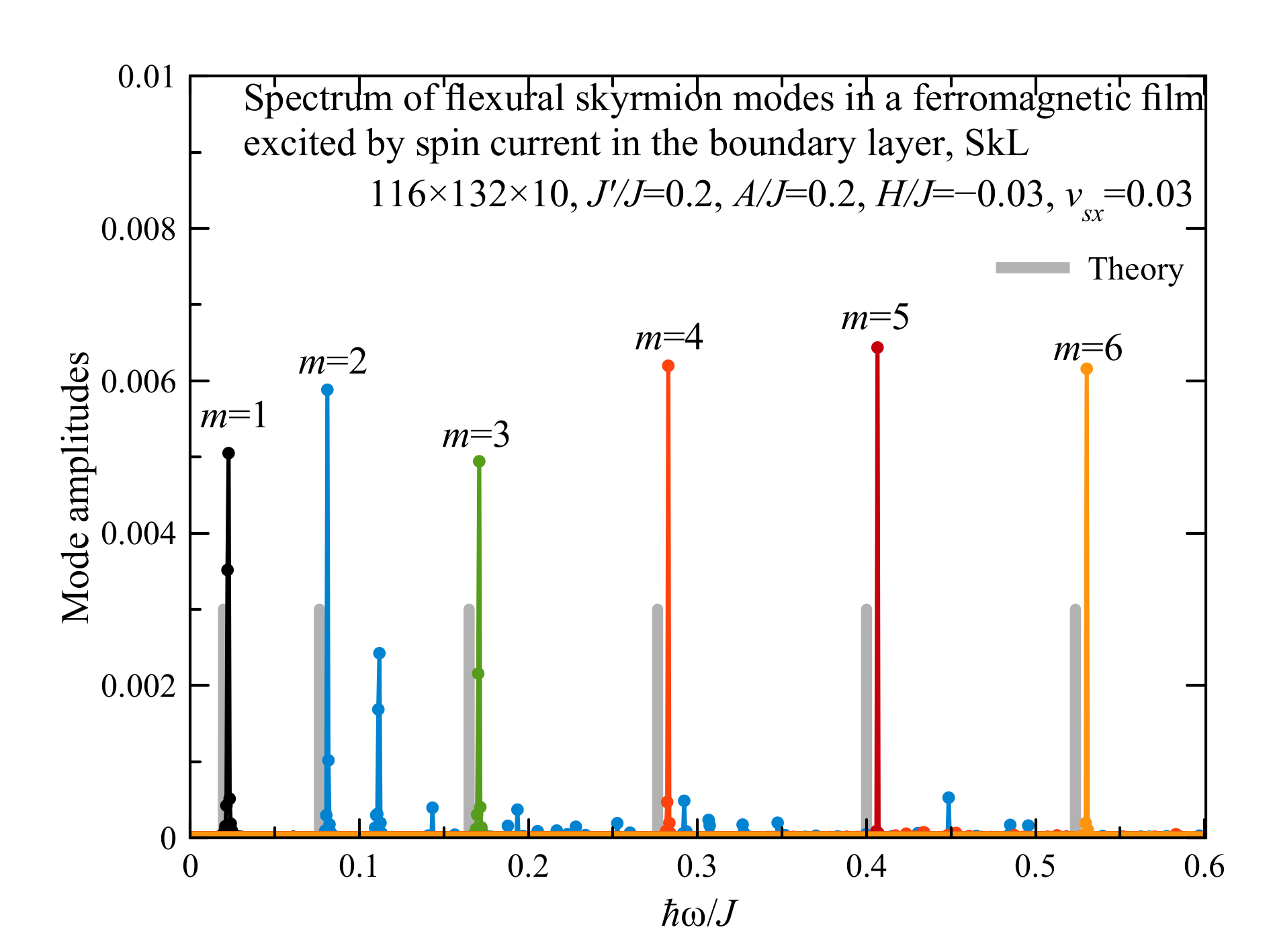} 
\par\end{centering}
\caption{Spectrum of the flexural modes in a lattice of skyrmion strings computed
using projection onto the flexural eigenvalues, Eq. (\ref{Rm_def}).
Here the noise is largely suppressed, and there is only one satellite
of the $m=2$ mode.}
\label{Fig_MA_SkL}
\end{figure}

\begin{figure}
\begin{centering}
\includegraphics[width=8cm]{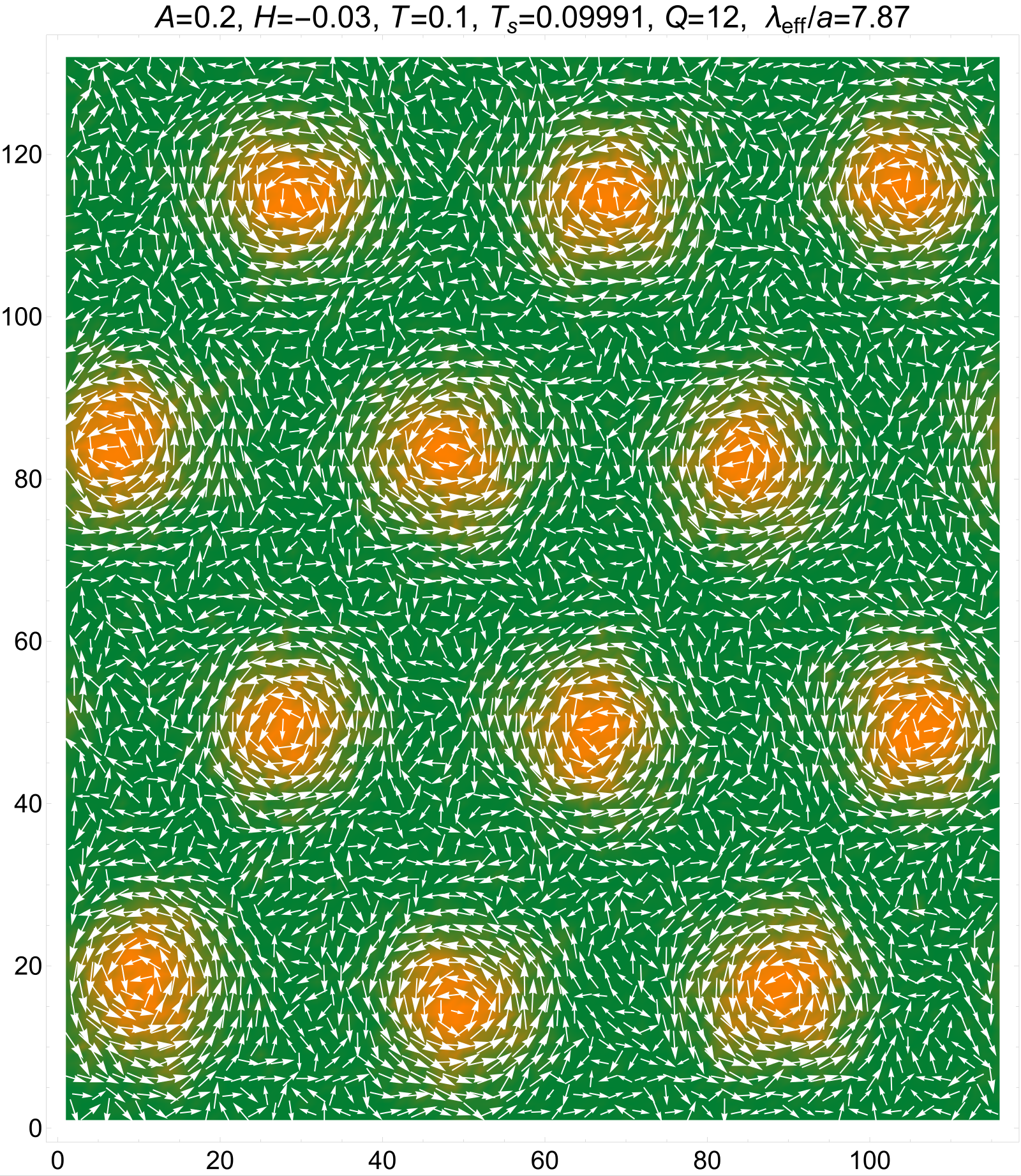} 
\par\end{centering}
\caption{Thermal spin configuration in the first layer of the $116\times132\times10$
system with an SkL of skyrmion strings at $T/J=0.1$. }
\label{Fig_Thermal-spin-configuration}
\end{figure}

\begin{figure}
\begin{centering}
\includegraphics[width=8cm]{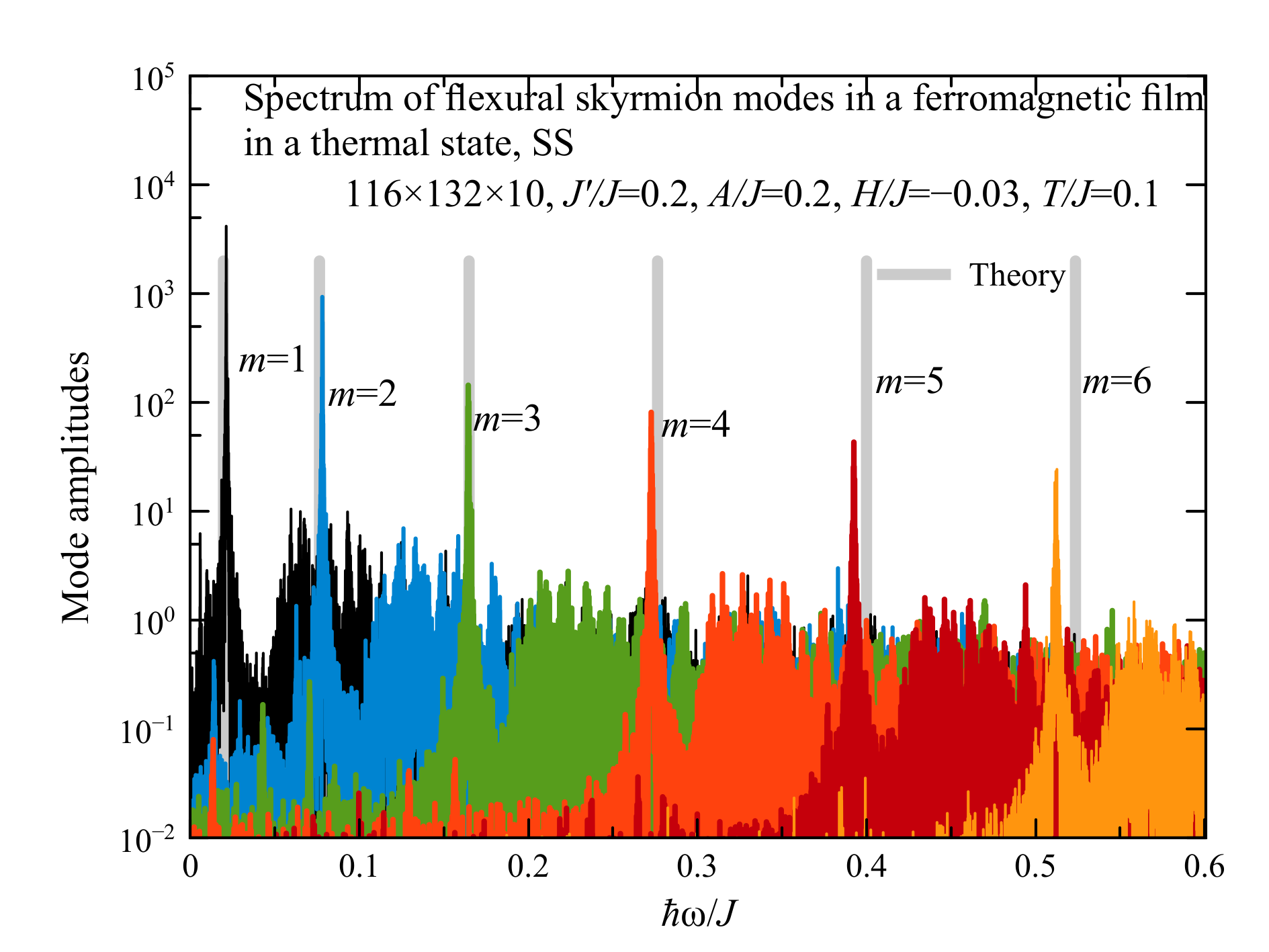} 
\par\end{centering}
\caption{Spectrum of the flexural modes of a single skyrmion string at $T/J=0.1$. }
\label{Fig_SS_T=00003D00003D0.1}
\end{figure}

\begin{figure}
\begin{centering}
\includegraphics[width=8cm]{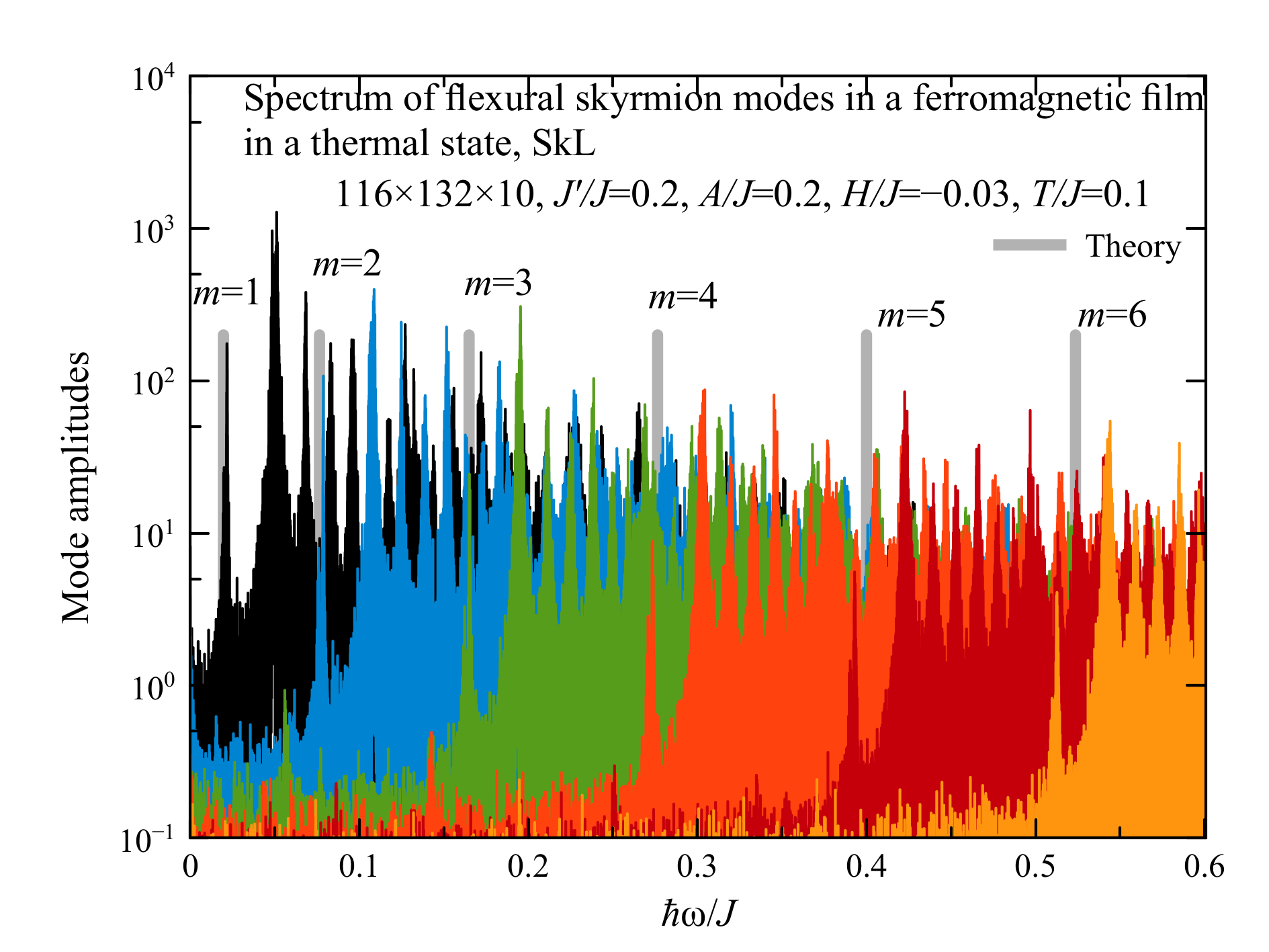} 
\par\end{centering}
\caption{Spectrum of excitations of an SkL of skyrmion strings at $T/J=0.1$.
Here the flexural modes with $m>3$ become indistinguishable from
the background.}
\label{Fig_SkL_T=00003D00003D0.1. }
\end{figure}

The spectrum of flexural oscillations in a single skyrmion string
obtained by solving the Landau-Lifshitz equation with a sinc spin
current in the boundary layer for a $50\times50\times10$ system and
computing the fluctuation spectrum of the $X$ component of the displacement
of the skyrmion in the boundary layer is shown in Fig. \ref{Fig_FSX_SS}.
In this case, the skyrmion size computed with the help of Eq. (\ref{lam_eff_def})
is $\lambda_{\mathrm{eff}}/a=8.64.$ There is a good agreement with
the analytical result for the modes' frequencies, Eq. (\ref{string-general}).
The numerically computed frequencies of the modes (peaks of the FS)
are only weakly dependent on the anisotropy $D$ and the applied field
$H$.

A similar computation for a larger $116\times132\times10$ system
containing an SkL of 12 skyrmion strings ($\lambda_{\mathrm{eff}}/a\approx7.98$)
shows a noisy result with FS peaks in accordance with the theory accompanied
by apparently random peaks, which are different in computations performed
on different machines, see Fig. \ref{Fig_FSx-SkL}. This result suggests
projecting on the modes' eigenfunctions, Eq. (\ref{Rm_def}), to suppress
the noise.

Figure \ref{Fig_MA_SkL} obtained by using Eq. (\ref{Rm_def}) is
indeed much cleaner than Fig. \ref{Fig_FSx-SkL} and demonstrates
a good agreement between the analytical theory and numerical results.

Let us now investigate flexural modes in thermal states. We choose
the temperature $T/J=0.1$, which is much lower than the Curie temperature
$T_{C}\sim J$. A typical SkL spin configuration obtained by Monte
Carlo is shown in Fig. \ref{Fig_Thermal-spin-configuration}. There
are 12 skyrmion strings, and the corresponding value of the topological
charge is $Q=12$. The spin temperature $T_{s}$ is very close to
the set temperature $T$. The energy per spin of this state at $T=0$
obtained by energy minimization is $E=0.069731,$which is below the
energy of the uniform state $E_{\mathrm{uni}}=0.13$.

Fluctuation spectra of the flexural modes of a single skyrmion (SS)
string and of an SkL of skyrmion strings in a thermal state are shown
in Figs. \ref{Fig_SS_T=00003D00003D0.1} and \ref{Fig_SkL_T=00003D00003D0.1. }.
For the SS, FS has peaks at the positions close to the theoretical
prediction, Eq. (\ref{string-general}). Thermal broadening of the
peaks is asymmetric, with more intensity on the right side. For the
SkL, the picture is more complicated as there are higher peaks to
the right of the theoretical positions, together with lower peaks
near the theoretical positions. This suggests a strong hybridization
of SkL's flexural modes with other SkL modes.

\section{Discussion}

\label{Sec_discussion}

We have studied flexural oscillations of skyrmion strings in a ferromagnetic
film consisting of $N$ atomic layers. Rigorous analytical result
for the frequencies of oscillations reads $|\omega_{m}|=(4J'S/\hbar)\sin^{2}[\pi m/(2N)]$,
where $J'$ is the interlayer exchange coupling, $S$ is the length
of the atomic spin, and $m=1,...,N-1$. It applies to conventional
ferromagnets possessing skyrmions, as well as to synthetic multilayers.
The latter permit control of the frequency range by controlling the
interlayer exchange constant $J'$. We have shown that these oscillations
can be excited by a spin torque exerted on the end of the skyrmion
string by the oscillating current in a (heavy metal) substrate.

Analytical results have been tested numerically in a microscopic 3D
ferromagnetic model with Dzyaloshinskii-Moriya interaction that supports
skyrmions. Individual skyrmion strings and lattices of skyrmion strings
have been studied in a system of over $1.5\times10^{5}$ spins at
finite temperature. The numerical method identifies the location of
skyrmion centers in 2D atomic layers, runs their dynamical evolution,
and computes the fluctuation spectrum at a fixed temperature.

For an isolated skyrmion string, the numerically obtained resonances
closely follow the analytical spectrum over the entire frequency and
temperature range studied. For a skyrmion-string lattice, the agreement
with analytically computed frequencies is good at low temperature.
At elevated temperatures, only the lowest flexural resonances remain
clearly identifiable, while higher modes are obscured by hybridization
with lattice and spin-wave excitations.

Flexural modes must provide an additional, interface-addressable channel
of microwave absorption. Their principal advantages over conventional
perpendicular standing spin waves are their direct coupling to a surface
spin-orbit torque, their potentially lower and denser resonance spectrum,
and their tunability through the skyrmion phase and skyrmion density.
Whether they dominate the total absorption depends on the spin Hall
conversion efficiency, interfacial transparency, damping, skyrmion
density, and competition with conventional ferromagnetic and spin-wave
resonances.

\section{Acknowledgements}

This work has been supported by Grant No. FA9550-24-1-0090 funded
by the Air Force Office of Scientific Research.\\


\begin{thebibliography}{10}
\bibitem{Bogdanov-NatPhy2020} A. N. Bogdanov and C. Panagopoulos,
Physical foundations and basic properties of magnetic skyrmions, Nature
Reviews Physics \textbf{2}, 492-498 (2020).

\bibitem{Bogdanov1989} A. N. Bogdanov and D. A. Yablonskii, Thermodynamically
stable ``vortices'' in magnetically ordered crystals. The mixed
state of magnets. Soviet Physics JETP \textbf{68}, 101-103 (1989).

\bibitem{Bogdanov94} A. Bogdanov and A. Hubert, Thermodynamically
stable magnetic vortex states in magnetic crystals, J. Magn. Magn.
Mater. \textbf{138}, 255-269 (1994).

\bibitem{Bogdanov-Nature2006} U. K. R$\ddot{\text{o}}${ß}ler,
N. Bogdanov, and C. Pfleiderer, Spontaneous skyrmion ground states
in magnetic metals, Nature \textbf{442}, 797-801 (2006).

\bibitem{Heinze-Nature2011} S. Heinze, K. von Bergmann, M. Menzel,
J. Brede, A. Kubetzka, R. Wiesendanger, G. Bihlmayer, and S. Blügel,
Spontaneous atomic-scale magnetic skyrmion lattice in two dimensions,
Nature Physics \textbf{7}, 713-718 (2011).

\bibitem{Boulle-NatNano2016} O. Boulle, J. Vogel, H. Yang, S. Pizzini,
D. de Souza Chaves, A. Locatelli, T. O. Mentes, A. Sala, L. D. Buda-Prejbeanu,
O. Klein, M. Belmeguenai, Y. Roussigné, A. Stahkevich, S. M. Chérif,
L. Aballe, M. Foerster, M. Chshiev, S. Auffret, I. M. Miron, and G.
Gaudin, Room-temperature chiral magnetic skyrmions in ultrathin magnetic
nanostructures, Nature Nanotechnology \textbf{11}, 449-454 (2016).

\bibitem{Leonov-NJP2016} A. O. Leonov, T. L. Monchesky, N. Romming,
A. Kubetzka, A. N. Bogdanov, and R. Wiesendanger, The properties of
isolated chiral skyrmions in thin magnetic films, New Journal of Physics
\textbf{18}, 065003-(16) (2016).

\bibitem{Leonov-NatCom2015} A. O. Leonov and M. Mostovoy, Multiply
periodic states and isolated skyrmions in an anisotropic frustrated
magnet, Nature Communications \textbf{6}, 8275-(8) (2015).

\bibitem{Zhang-NatCom2017} X. Zhang, J. Xia, Y. Zhou, X. Liu, H.
Zhang, and M. Ezawa, Skyrmion dynamics in a frustrated ferromagnetic
film and current-induced helicity locking-unlocking transition, Nature
Communications \textbf{8} 1717-(10) (2017).

\bibitem{IvanovPRB06} B. A. Ivanov, A. Y. Merkulov, V. A. Stepanovich,
C. E. Zaspel, Finite energy solitons in highly anisotropic two dimensional
ferromagnets, Physical Review B \textbf{74}, 224422-(17) (2006).

\bibitem{Lin-PRB2016} S.-Z. Lin and S. Hayami, Ginzburg-Landau theory
for skyrmions in inversion-symmetric magnets with competing interactions,
Physical Review B \textbf{93}, 064430-(16) (2016).

\bibitem{CG-NJP2018} E. M. Chudnovsky and D. A. Garanin, Skyrmion
glass in a 2D Heisenberg ferromagnet with quenched disorder, New Journal
of Physics \textbf{20} 033006-(9) (2018).

\bibitem{Moutafis-PRB2009} C. Moutafis, S. Komineas, and J. A. C.
Bland, Dynamics and switching processes for magnetic bubbles in nanoelements,
Physical Review B \textbf{79}, 224429-(8) (2009).

\bibitem{Fert-NatMat2017} A. Fert, N. Reyren, and V. Cros, Magnetic
skyrmions: advances in physics and potential applications. Nature
Review Materials \textbf{2}, 17031-(17) (2017).

\bibitem{APL-2021} S. Luo and L. You, Skyrmion devices for memory
and logic applications, APL Materials \textbf{9}, 050901 (2021).

\bibitem{Mochizuki-PRL2012} M. Mochizuki, Spin-wave modes and their
intense excitation effects in skyrmion crystals, Physical Review Letters
\textbf{108}, 017601-(5) (2012).

\bibitem{Onose-PRL2012} Y. Onose,Y. Okamura, S. Seki, S. Ishiwata,
and Y. Tokura, Observation of magnetic excitations of skyrmion crystal
in a helimagnetic insulator Cu$_{2}$OSeO$_{3}$, Physical Review
Letters \textbf{109}, 037603-5 (2012).

\bibitem{DA-RJ-EC-PRB2020} D. A. Garanin, R. Jaafar, and E. M. Chudnovsky,
Breathing mode of a skyrmion on a lattice, Physical Review B \textbf{101},
014418 (2020).

\bibitem{Aqeel-PRL2021} A. Aqeel, J. Sahliger, T. Taniguchi, S. Mändl,
D. Mettus, H. Berger, A. Bauer, M. Garst, C. Pfleiderer, and C. H.
Back, Microwave Spectroscopy of the Low-Temperature Skyrmion State
in Cu$_{2}$OSeO$_{3}$, Physical Review Letters \textbf{126}, 017202-(7)
(2021).

\bibitem{Satywali-NatCom2021} B. Satywali, V. P. Kravchuk, L. Pan,
M. Raju, S. He, F. Ma, A. P. Petrovi\'{c}, M. Garst, and C. Panagopoulos,
Nature Communications \textbf{12}, 1909-(8) (2021).

\bibitem{Lee-JPhys2022} O. Lee, J. Sahliger, A. Aqeel , S. Khan,
S. Seki, H. Kurebayashi, and C. H. Back, Tunable gigahertz dynamics
of low-temperature skyrmion lattice in a chiral magnet, Journal of
Physics: Condensed Matter \textbf{34}, 095801-(10) (2022).

\bibitem{Li-JPhys2023} Y. Li, X. Wang, and L. Ma, Instability of
skyrmion lattice under microwave magnetic field due to single-q helimagnetic
excitation mode. Journal of Physics: Condensed Matter \textbf{35},
105801-(7) (2023).

\bibitem{DG-EC-PRB2025} D. A. Garanin and E. M. Chudnovsky, Skyrmion
crystal in a microwave field, Physical Review B \textbf{112}, 024422-(13)
(2025).

\bibitem{EC-DG-PRB2026} E. M. Chudnovsky and D. A. Garanin, Skyrmion
cyclotron resonance in ferromagnets, Physical Review B \textbf{113},
224414-(8) (2026).

\bibitem{DG-EC-MMM2026} D. A. Garanin and E. M. Chudnovsky, Massive
dynamics of skyrmions in ferrimagnetic films, Journal of Magnetism
and Magnetic Materials \textbf{654}, 174226-(13) (2026).

\bibitem{Yokouchi-SviAdv2018} T. Yokouchi, S. Hoshino, N. Kanasawa,
A. Kikkawa, D. Morikawa, K. Shibata, T.-h. Arima, Y. Taguchi, F. Kagawa,
N. Nagaosa, and Y. Tokura, Current-induced dynamics of skyrmion strings,
Science Advances \textbf{4}, 1115-(7) (2018).

\bibitem{Seki-NatCom2020} S. Seki, M. Garst, J. Waizner, R. Takagi,
N. D. Khanh, Y. Okamura, K. Kondou, F. Kagawa, Y. Otani, and Y. Tokura,
Propagation dynamics of spin excitations along skyrmion strings, Nature
Communications \textbf{11}, 256-(7) (2020).

\bibitem{Seki-NatMat2022} S. Seki, M. Suzuki, M. Ishibashi, R. Takagi,
N. D. Khanh, Y. Shiota, K. Shibata, W. Koshibae, Y. Tokura, and T.
Ono, Direct visualization of the three-dimensional shape of skyrmion
strings in a noncentrosymmetric magnet, Nature Materials \textbf{21},
181-187 (2022).

\bibitem{Kravchuk-PRB2023} V. P. Kravchuk, Nonlinear dynamics of
skyrmion strings, Physical Review B \textbf{108}, 144412-(18) (2023).

\bibitem{Ran-NatCom2024} K. Ran, W. Tan, X. Sun, Y. Liu, R. M. Dalgllesh,
N.-J. Steinke, G. van der Laan, S. Langridge, T. Hesjedal, and S.
Zhang, Bending skyrmion strings under two-dimensional thermal gradients,
Nature Communications \textbf{15}, 4860-8 (2024).

\bibitem{Panigrahy2022} S. Panigrahy, S. Mallick, J. Sampaio, and
S. Rohart, Skyrmion inertia in synthetic antiferromagnets, Physical
Review B \textbf{106}, 144405-(8) (2022).

\bibitem{Nowak-PRB2023} M. Weißenhofer and U. Nowak, Temperature
dependence of current-driven and Brownian skyrmion dynamics in ferrimagnets
with compensation point, Physical Review B \textbf{107}, 0644423-(9)
(2023).

\bibitem{Lau2025} M. Lau, W. Häusler, and M. Thorwart, Moving skyrmions
in antiferromagnets by sublattice displacements, Physical Review B
\textbf{111}, 144411-(10) (2025).

\bibitem{BP} A. A. Belavin and A. M. Polyakov, Metastable states
of two-dimensional isotropic ferromagnets, Pis'ma Zh. Eksp. Teor.
Fiz. \textbf{22}, 503-5-6 (1975) {[}JETP Letters \textbf{22}, 245-248
(1975){]}.

\bibitem{Thiele} A. A. Thiele, Steady-state motion of magnetic domains,
Physical Review Letters \textbf{30}, 230-233 (1973).

\bibitem{Strang-1999} G. Strang, The discrete cosine transform, SIAM
Review \textbf{41}, 135-147 (1999).

\bibitem{Gobel-2021} B. Göbel and I Mertig, Skyrmion ratchet propagation:
utilizing the skyrmion Hall effect in AC racetrack storage devices,
Nature Scientific Reports \textbf{11}, 3020-((11) (2021).

\bibitem{Yang-2023} S. Yang, Y. Zhao, K. Wu, Z. Chu, X. Xu, X. Li,
J. Akerman, and Y. Zhou, Reversible conversion between skyrmions and
skyrmioniums, Nature Communications \textbf{14}, 3406-(8) (2023).

\bibitem{gar25jpcm} D. A. Garanin, Energy minima and ordering in
ferromagnets with static randomness, Journal of Physics: Condensed
Matter, \textbf{37}, 385803 (2025).

\bibitem{garchu24jmmm} D. A. Garanin and E. M. Chudnovsky, Solid--liquid
transition in a skyrmion matter, Journal of Magnetism and Magnetic
Materials, \textbf{606}, 172395 (2024).

\bibitem{kacgar2001PA_surf} H. Kachkachi and D. A. Garanin, Boundary
and finite-size effects in small magnetic systems, Physics A, \textbf{300},
487 (2001).

\bibitem{caichugar2012prb} Liufei Cai, E. M. Chudnovsky, and D. A.
Garanin, Collapse of skyrmions in two-dimensional ferromagnets and
antiferromagnets, Physical Review B \textbf{86}, 024429 (2012).

\bibitem{eriksson90mm} F. Eriksson, On the Measure of Solid Angles,
Mathematics Magazine, \textbf{63}, 184-18 (1990).

\bibitem{gar21pre} D. A. Garanin, Energy balance and energy correction
in dynamics of classical spin systems, Physical Review E, \textbf{104},
055306 (2021).

\end{thebibliography}
\end{document}